\documentclass[12pt]{article}
\usepackage{amsmath,amssymb}
\usepackage{graphicx}

\newcommand{\Real}{\mathop\mathrm{Re}\nolimits}
\newcommand{\Imag}{\mathop\mathrm{Im}\nolimits}

\newcommand{\cL}{\mathcal{L}}

\newcommand{\bra}[1]{\langle #1 |}
\newcommand{\ket}[1]{| #1 \rangle}

\newcommand{\e}{\mathrm{e}}

\newcommand{\sm}[1]{{\scriptscriptstyle \rm #1}}

\renewcommand{\l}{\left(}
\renewcommand{\r}{\right)}
\newcommand{\la}{\langle}
\newcommand{\ra}{\rangle}  

\providecommand{\href}[2]{#2}

\begin{document}

\title{T-Odd Correlations in $\pi\to e\nu_e\gamma$ and
  $\pi\to\mu\nu_\mu\gamma$ Decays}

\author{
  F.~L.~Bezrukov\thanks{Email: \texttt{fedor@ms2.inr.ac.ru}}\;, 
  D.~S.~Gorbunov\thanks{Email: \texttt{gorby@ms2.inr.ac.ru}}\\
  {\small{\em
      Institute for Nuclear Research of the Russian Academy of Sciences, }}\\
  {\small{\em
      60th October Anniversary prospect 7a, Moscow 117312, Russia
    }}
}
\date{}
\maketitle

\begin{abstract}
  The transverse lepton polarization asymmetry in $\pi_{l2\gamma}$
  decays may probe T-violating interactions beyond the Standard Model.
  Dalitz plot distributions of the expected
  effects are presented and
  compared to the contribution from the Standard Model final state
  interactions.  We 
  give an example 
  of a phenomenologically viable model, where a considerable
  contribution to the transverse lepton polarization asymmetry
  arises.  
\end{abstract}

\section{Introduction}

T-violation beyond the Standard Model is usually searched for in
decays forbidden by time reversal symmetry.  Another way to probe
T-violation are measurements of T-odd observables in allowed decays of
mesons.  The well known example is $K^0\to\pi^+\pi^-e^+e^-$ decay,
where T-odd correlation is experimentally observed
\cite{Alav-Harati:1999ff} and coincides with theoretical prediction of
the Standard Model \cite{Sehgal:1992wm}.  Other widely considered
T-odd observables are transverse muon polarizations ($P_T$) in
$K\to\pi\mu\nu$ and $K\to\mu\nu\gamma$ decays.  There is no tree level
SM contribution to $P_T$ in these decays, so they are of a special
interest for search for new physics.  Unfortunately, $P_T$ is not
exactly zero in these decays even in T-invariant theory~---
electromagnetic loop corrections contribute to $P_T$ and should be
considered as a background.  There are no experimental evidence for
nonzero $P_T$ in these processes at present time
\cite{Abe:1999nc,Kudenko:2000sc}, but the sensitivity of the
experiments has not yet reached the level of SM loop contributions.

In this paper we study the decays $\pi\to e\nu\gamma$ and
$\pi\to\mu\nu\gamma$.  Within the Standard Model, 
T-violation in these processes does not appear
at tree level, but interactions, contributing to
it, emerge in various extensions of SM.  We shall demonstrate that
$\pi_{l2\gamma}$ decays are attractive probes of new physics beyond
the Standard Model.  Depending on the model, $\pi_{l2\gamma}$ decays
may be even more attractive than usually considered
$K_{l2\gamma}$ decays.  

Though $\pi_{e2\gamma}$ decay has very small branching ratio (it is of
order $10^{-7}$), we find that the distribution of the transverse
electron polarization over the Dalitz plot significantly overlaps with
the distribution of differential branching ratio, as opposed to the
situation with $K_{\mu2\gamma}$ decay.  Moreover, the contribution of
FSI (final-state interactions related to SM one-loop diagrams) to the
observable asymmetry, being at the level of $10^{-3}$, becomes even
smaller in that region of the Dalitz plot, where the contribution from
new effective T-violating interaction is maximal. Thus,
$\pi_{e2\gamma}$ decay is potentially quite interesting probe of
T-violation, though it is worth noting that the experimental
measurement of electron or positron polarization is quite complicated.

The $\pi_{\mu2\gamma}$ decay has much higher branching ratio than
$\pi_{e2\gamma}$ decay, and experimental measurement of muon
polarization is simpler than that of electron.  Unlike in
$\pi_{e2\gamma}$ decay, however, FSI contributions and contributions from
T-violating interaction to the transverse muon polarization are of
similar shape, as we show in this paper.  
This means that only those new physics effects may be detected, 
which are stronger 
than FSI, but if this is the case then detecting
T-violation in $\pi_{\mu2\gamma}$ needs much less pion statistics,
than that required in the case of $\pi_{e2\gamma}$.

To demonstrate that pion decays may be relevant processes where the
signal of new physics may be searched for, we present a simple model
of heavy pseudoscalar particle exchange leading in the low energy
limit to the T-violating four fermion interaction.  We find the
constraints on the parameters of this model coming from various other
experiments and describe regions of the parameter space which result
in large T-violating effects in $\pi_{l2\gamma}$ decays.  Depending on
the parameters of the model, an experiment measuring transverse lepton
polarization with pion statistics of $10^5\div10^{10}$ pions for
$\pi_{\mu2\gamma}$ decay and $10^8\div10^{13}$ pions for
$\pi_{e2\gamma}$ decay is needed to detect the T-violating effects
(taking into account statistical uncertainty only and assuming ideal
experimental efficiencies).

The paper is organized as follows.  In section \ref{sec:2} we
introduce a generic effective four-fermion interaction giving rise to
T-odd correlation of lepton transverse polarization and calculate the
distributions of this polarization and of the differential branching
ratio of $\pi_{l2\gamma}$ decay over the Dalitz plot.  In section
\ref{sec:2.1} we estimate the contribution of the final state
interactions to the observable asymmetry.  Section \ref{sec:2.2} is
devoted to the constraint on the effective Lagrangian coming from the
measurement of $\pi\to l\nu_l$ decays. Generically, this constraint is
quite strong, but it becomes much less restrictive if there is 
a hierarchy of the constants in the Lagrangian responsible for the
observable T-violating effects.  An example of a high 
energy model of T-violation 
is presented in section \ref{sec:3}.  The constraints on
the parameters of this model emerging from the measurements of muon
life-time, from the study of parameters of kaon mixing and from the
experimental limit on neutron dipole moment are considered in sections
\ref{sec:3.1}, \ref{sec:3.2} and \ref{sec:3.3}, respectively.  It is
shown that for generic values of the model parameters, the 
measurement of CP-violating parameter $\epsilon$ forbids 
new observable T-violating effects in $\pi_{l2\gamma}$ decays, 
but if the parameters exhibit a
special pattern, the T-violating effects in $\pi_{l2\gamma}$ may be large.

\section{T-violating effect in $\pi^+_{l2\gamma}$ decay}
\label{sec:2}

Let us consider the simplest effective four-fermion
interaction\footnote{Additional scalar, vector- or axial-type
  interactions do not contribute to lepton transverse 
polarization~\cite{Chen:1997gf}.} 
\begin{equation}\label{4-fermion}
  {\cal L}_{\mathrm{eff}} =
  G^l_{\sm P}\bar{d}\gamma_5u\cdot\bar{\nu}_l(1+\gamma_5)l
  +\mathrm{h.c}.
\end{equation}
that may be responsible for the T-violating effects in pion physics
beyond the Standard Model.  Indeed, the imaginary part of the constant
$G^l_{\sm P}$ contributes to transverse lepton polarization in decays
$\pi^+_{l2\gamma}$.  

To calculate this polarization let us write the amplitude of the decay
$\pi^+_{l2\gamma}$ in terms of inner brems\-strah\-lung (IB) and
structure-dependent (SD)
contributions~\cite{Gabrielli:1993dp,Bijnens:1993en,Chen:1997gf}
\begin{equation*}
  M=M_\mathrm{IB}+M_\mathrm{SD} \;,
\end{equation*}
with
\begin{equation}
  M_\mathrm{IB} = ie\frac{G_F}{\sqrt{2}}\cos\theta_c f_\pi m_l
    \epsilon^*_\alpha K^\alpha \;,\qquad
  M_\mathrm{SD} = -ie\frac{G_F}{\sqrt{2}}\cos\theta_c
    \epsilon^*_\mu L_\nu H^{\mu\nu} \;,
\end{equation}
where
\begin{align}
  K^\alpha &=
    \bar{u}(k)(1+\gamma_5) \left(
      \frac{p^\alpha}{pq}
      -\frac{2p^\alpha_l+\hat{q}\gamma^\alpha}{2p_lq}
    \right) v(p_l,s_l) \;,\nonumber\\
  L_\nu &=
    \bar{u}(k)\gamma_\nu(1-\gamma_5)v(p_l,s_l) \;,\nonumber\\
  H^{\mu\nu} &=
    \frac{F_A}{m_\pi}(-g^{\mu\nu}pq+p^\mu q^\nu)
    +i\frac{F_V}{m_\pi}\varepsilon^{\mu\nu\alpha\beta}q_\alpha p_\beta \;.
  \label{Fvdef}
\end{align}
with convention for Levi-Civita tensor $\varepsilon^{0123}=-1$.  Here
$\epsilon_\alpha$ is the photon polarization vector, $p$, $k$, $p_l$,
$q$ are the four-momenta of $\pi^+$, $\nu_l$, $l^+$, and $\gamma$,
respectively, $s_l$ is the polarization vector of the charged lepton, $F_V$,
$F_A$ are vector and axial form factors of pion radiative decay; the
effect coming from the Lagrangian \eqref{4-fermion} may be absorbed 
in the $f_\pi$ form factor
\begin{equation}\label{fpi}
  f_\pi = f^0_\pi(1+\Delta^l_P) \;,\qquad f^0_\pi\approx130.7~\mathrm{MeV}
\end{equation}
with
\begin{equation*}
  \Delta^l_P = \frac{\sqrt{2}G^l_P}{G_F\cos\theta_c}\cdot
    \frac{B_0}{m_l}\;,\qquad B_0=-\frac{2}{(f^0_\pi)^2}\la 0|\bar{q}q|0\ra=
  \frac{m_\pi^2}{m_u+m_d}\approx2\mathrm{GeV}\;.  
\end{equation*}

We write the components of $s_l$ in terms of $\vec{\xi}$, a unit
vector along the lepton spin in its rest frame, as follows, 
\begin{align*}
  s_0 &= \frac{\vec{\xi}\vec{p}_l}{m_l} \;,
  &
  \vec{s} &= \vec{\xi}+ \frac{s_0}{E_l+m_l}\vec{p}_l \;.
\end{align*}
In the rest frame of $\pi^+$, the partial decay width into the state
with lepton spin $\vec{\xi}$ is found to be
\begin{equation*}
  d\Gamma(\vec{\xi}\,)=\frac{1}{2m_\pi }|M|^2 (2\pi)^4\delta(p-p_e-k-q)
  \frac{d\vec{q}}{2E_q(2\pi)^3} 
  \frac{d\vec{p}_l}{2E_l(2\pi)^3}
  \frac{d\vec{k}}{2E_\nu(2\pi)^3} \;,
\end{equation*}
with
\begin{equation*}
  |M|^2=\rho_0(x,y)\left[1 + 
(P_L\vec{e}_L +P_N\vec{e}_N + P_T\vec{e}_T)\cdot\vec{\xi}~\!\right] \;,
\end{equation*}
where $\vec{e}_i\ (i=L,N,T)$ are the unit vectors along the 
longitudinal ($P_L$), normal ($P_N$) and transverse ($P_T$) components
of the lepton polarization, defined by
\begin{equation}
  \vec{e}_L= \frac{\vec{p}_l }{ |\vec{p}_l|} \;,\qquad
  \vec{e}_N= \frac{\vec{p}_l\times (\vec{q}\times \vec{p}_l)
    }{ |\vec{p}_l\times (\vec{q}\times \vec{p}_l)|} \;,\qquad
  \vec{e}_T = \frac{\vec{q}\times \vec{p}_l}{
    |\vec{q}\times \vec{p}_l|} \;,
\end{equation} 
respectively, and
\begin{equation*}\begin{split}
  \rho_0(x,y) &= e^2 \frac{G^2_F}{2} \cos^2\!\theta_c 
    (1-\lambda)\left\{ \frac{4m^2_l|f_\pi|^2
    }{ \lambda x^2}\left[x^2+2(1-r_l)\left(1-x-\frac{r_l}{
    \lambda}\right)\right]\right.
  \\
  &\quad + m^4_\pi x^2\left[|F_V+F_A|^2\frac{\lambda^2 }{ 1-\lambda}
    \left(1-x-\frac{r_l}{ \lambda}\right)+
    |F_V-F_A|^2(y-\lambda)\right]
  \\
  &\quad - 4m_\pi m^2_l\left[\Real[f_\pi(F_V+F_A)^*]\left(1-x-\frac{r_l}{
    \lambda}\right) \right.
  \\
  &\quad - \left.  \left.  \Real[f_\pi(F_V-F_A)^*]\frac{1-y+\lambda }{ 
    \lambda}\right]\right\} \;,
\end{split}\end{equation*}
with $\lambda=(x+y-1-r_l)/x$, $r_l=m^2_l/M^2_\pi$, and
$x=2pq/p^2=2E_{\gamma}/m_\pi$ and $y=2pp_l/p^2=2E_l/m_\pi$ are
normalized energies of the photon and charged lepton, respectively.  In
terms of these variables the differential decay width reads 
\begin{equation}\label{dgamma}
  d\Gamma(\vec{\xi}\,) = \frac{m_\pi}{32(2\pi)^3}|M(x,y,\vec{\xi}\,)|^2 dx\, dy
          = \frac{m_\pi}{32(2\pi)^3}|M(x,\lambda,\vec{\xi}\,)|^2 x dx\, d\lambda
\end{equation}
For the transverse lepton polarization asymmetry 
\begin{equation}\label{Ptdef}
  P_T(x,y)=\frac{d\Gamma(\vec{e}_T)-d\Gamma(-\vec{e}_T) }{
  d\Gamma(\vec{e}_T)+d\Gamma(-\vec{e}_T)} \,, 
\end{equation}
we find 
\begin{equation*}
  P_T(x,y)=\frac{\rho_T(x,y)}{ \rho_0(x,y)}\,, 
\end{equation*}
with
\begin{equation*}\begin{split}
  \rho_T(x,y) &= -2e^2G^2_F \cos^2\!\theta_c m^2_\pi m_l
    \frac{1-\lambda }{ \lambda}
    \sqrt{\lambda y-\lambda^2-r_l } 
  \\
  &\quad \times\left\{ \Imag[f_\pi (F_V+F_A)^*]\frac{\lambda}{1-\lambda}
    \left(1-x-\frac{r_l}{\lambda}\right) +
    \Imag[f_\pi (F_V-F_A)^*] \right\} \;.
\end{split}\end{equation*}
The asymmetry $P_T$ is odd under time-reversal (the sign of
$\vec{\xi}\vec{e}_T$ obviously changes under time-reversal), and $P_N$
and $P_L$, defined analogously to \eqref{Ptdef}, are even under time
reversal.  Moreover, one can show that interaction \eqref{4-fermion}
does not contribute to $P_N$ and $P_L$ \cite{Chen:1997gf}.  One
observes that $P_T$ is proportional to imaginary part of $\Delta_P$, so
it is convenient to rewrite Eq.~(\ref{Ptdef}) in the following form
\begin{equation}\label{pt}
  P_T(x,y)=[\sigma_V(x,y)-\sigma_A(x,y)]\cdot\Imag[\Delta^l_P]\,,
\end{equation}
where
\begin{gather*}
  \sigma_V(x,y)=2e^2G^2_F \cos^2\!\theta_c m^2_\pi  m_l f^0_\pi F_V
  \frac{\sqrt{\lambda y-\lambda^2-r_l }}{ \rho_0(x,y)}
  \left(\frac{\lambda -1}{ \lambda}-
  \left(1-x-\frac{r_l }{ \lambda}\right)\!\right),
  \\
  \sigma_A(x,y)=2e^2G^2_F \cos^2\!\theta_c m^2_\pi  m_l f^0_\pi F_A
  \frac{\sqrt{\lambda y-\lambda^2-r_l }}{ \rho_0(x,y)}
  \left(\frac{\lambda -1}{ \lambda}+
  \left(1-x-\frac{r_l }{ \lambda}\right)\!\right).
\end{gather*}

\begin{figure}[tb]
  \begin{center}
    \includegraphics{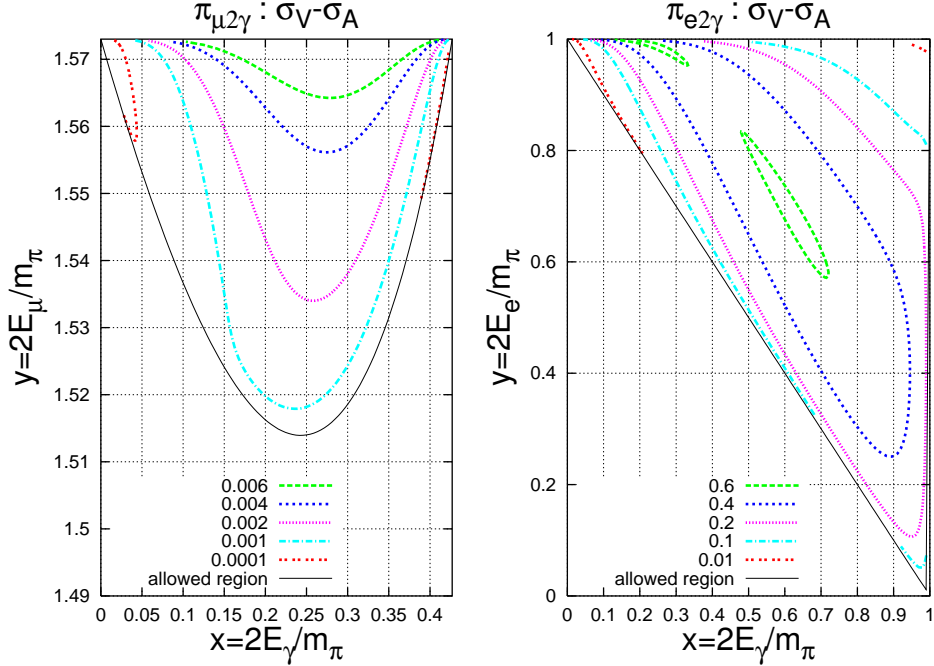}
  \end{center}
  \caption{The contour-plots for the function 
    $[\sigma_V-\sigma_A]$ for $\pi_{\mu2\gamma}$ and $\pi_{e2\gamma}$ 
     decays, which describe $P_T$ distribution over the Dalitz plot
    (see Eq.~(\ref{pt})).
  \label{fig:sigma}}
\end{figure}

Taking $F_V=-0.0259$ (CVC prediction) and 
$F_A=-0.0116$~\cite{pdg2000,Bolotov:1990yq} we present 
in Figure~\ref{fig:sigma} 
the contour-plot of 
$[\sigma_V-\sigma_A]$ as a function of $x$ and $y$ for
$\pi_{e2\gamma}$ and $\pi_{\mu2\gamma}$ decays.  

\begin{figure}[tb]
  \begin{center}
    \includegraphics{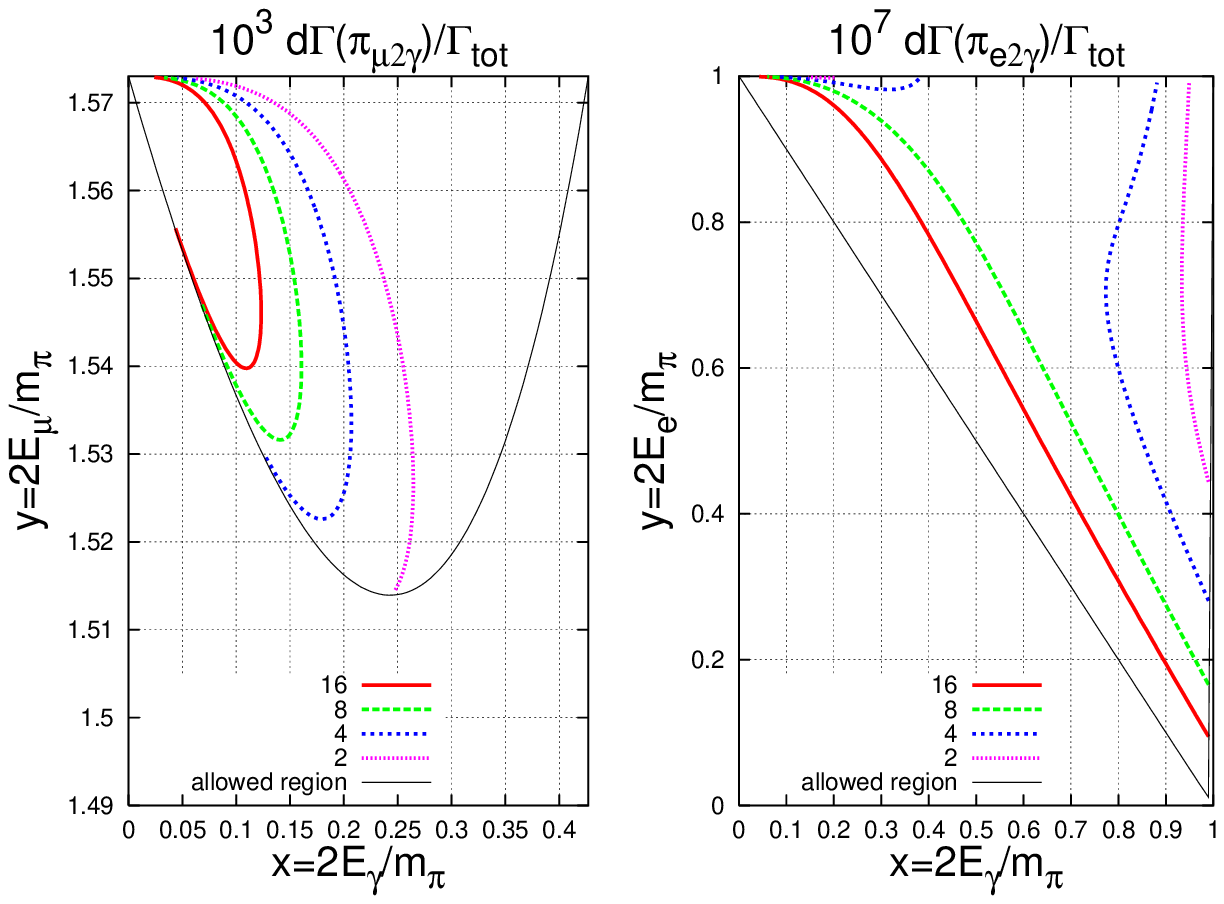}
  \end{center}
  \caption{The distribution of differential branching ratios 
    of $\pi_{\mu2\gamma}$ and $\pi_{e2\gamma}$ decays (multiplied by
    $10^3$ and $10^7$, respectively) over the Dalitz plot.
  \label{fig:dgamma}}
\end{figure}

As one can see, in a
large region of kinematic variables, 
$[\sigma_V-\sigma_A]$ is about 
$0.5$ for the $\pi_{e2\gamma}$ decay.  This means that transverse
electron polarization $P_T$ for this process,
Eq.~\eqref{pt}, is of the same
order as $\Imag[\Delta_P^e]\simeq5\times10^3\cdot\Imag[G_P^e/G_F]$.  It
is worth noting that the region of
the Dalitz plot where large
T-violating effect might be observed, significantly overlaps with the
region where the partial decay width $\Gamma(\pi_{e2\gamma})$ is
saturated (cf.\ Figures~\ref{fig:sigma} and \ref{fig:dgamma}).
This is in contrast to the situation with T-violation in
$K_{\mu2\gamma}$ decay (see, e.g., Ref.~\cite{Chen:1997gf}), where the
analogous overlap is small, so the differential branching ratio in the
relevant region is smaller than on average.

The situation with $\pi_{\mu2\gamma}$ decay is less attractive:
one finds that $\Imag[\Delta_P^\mu]\simeq25\cdot\Imag[G_P^\mu/G_F]$ and 
$[\sigma_V-\sigma_A]$ is of the order of $10^{-3}\div10^{-4}$ over the
Dalitz plot.  Moreover, the
decay rate is saturated in the region of small $x$ (low energy
photons), where the T-violating effect is small.  On the other hand,
the branching ratio is larger by three orders of magnitude than that
of $\pi_{e2\gamma}$ decay.

\subsection{Final-state interactions}
\label{sec:2.1}

\begin{figure}[tb]
  \begin{center}
    \includegraphics{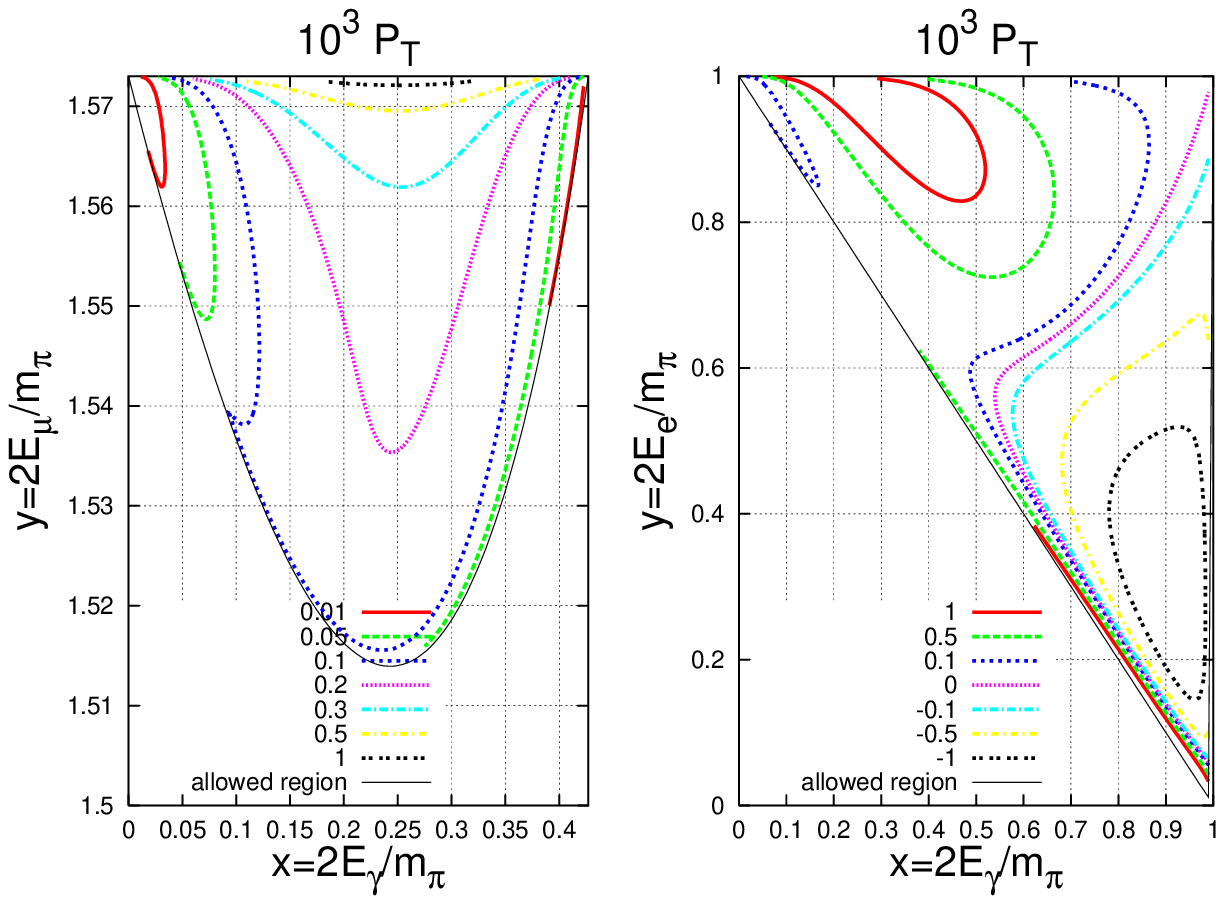}
  \end{center}
  \caption{Transverse lepton polarization due to FSI in
    $\pi_{\mu2\gamma}$ and $\pi_{e2\gamma}$ decays.
\label{fig:fsi}}
\end{figure}

Now let us estimate SM contribution to the T-violating observable
$P_T$.  This contribution arises due to final-state interactions (FSI)
--- one-loop diagrams with virtual photons.  These diagrams are
similar to the diagrams leading to FSI contribution in
$K\to\mu\nu_\mu\gamma$ decay.  The latter contribution was calculated in
Ref.~\cite{Braguta:2001ea}.  Thus FSI in $\pi_{l2\gamma}$ may be
estimated by making use of corresponding replacements ($m_K\to m_\pi$,
etc.) in formulae of Ref.~\cite{Braguta:2001ea}.  The
result is presented in Figure~\ref{fig:fsi}.

For the $\pi_{e2\gamma}$ decay, the ($x,y$)-distributions of FSI
contribution and the contribution from the four-fermionic
interaction~(\ref{4-fermion}) differ in shape.  Specifically, part of
the region with maximal $P_T$ from four-fermion interaction
\eqref{4-fermion} corresponds to the region of small $P_T$ from FSI\@.
This implies that if measured, $P_T$ distribution could probe
T-violating interaction~(\ref{4-fermion}) with an accuracy higher than
$\Imag[\Delta^e_P]\sim 10^{-3}$
($\Imag[G_P^e/G_F]\sim2\times10^{-7}$).  Again, this is not the case
for $K_{\mu2\gamma}$ decay (see Ref.~\cite{Braguta:2001ea}).

For the $\pi_{\mu2\gamma}$ decay the situation is less promising.  The
contributions from the lagrangian \eqref{4-fermion} and from FSI are
distributed over the Dalitz plot similarly.  So one may hope to probe
T-violating interaction only at the level
$\Imag[\Delta_P^\mu]\gtrsim10^{-1}$
($\Imag[G_P^\mu/G_F]\sim3\times10^{-3}$).

\subsection{Constraint from $\pi\to l\nu$ decays}
\label{sec:2.2}

The interaction term~(\ref{4-fermion}) not only gives rise to
T-violation in $\pi\to l\nu_l\gamma$ decays but also contributes to the
rate of $\pi\to l\nu_l$ decays.  Since the ratio of leptonic
decays of the pion has been accurately measured
\cite{pdg2000,Czapek:1993kc,Britton:1992pg,Bryman:1986bv},
\begin{equation}
  R=\frac{\Gamma(\pi\to e\nu) +\Gamma(\pi\to e\nu\gamma)}
         {\Gamma(\pi\to\mu\nu)+\Gamma(\pi\to\mu\nu\gamma)} =
         (1.230\pm0.004) \times 10^{-4}
  \;,
\label{R}
\end{equation}
the coupling constants $G_P^e$ and $G_P^\mu$ are strongly constrained.
Indeed, the standard $(V-A)$ theory of electroweak interactions
predicts
\begin{equation*}
  R_0 = \frac{m_e^2}{m_\mu^2}
        \frac{(m_\pi^2-m_e^2)^2}{(m_\pi^2-m_\mu^2)^2}
        (1+\delta)
      = 1.233 \times 10^{-4}
  \;,
\end{equation*}
where $\delta$ is the radiative correction
\cite{Berman:1958gx,Kinoshita:1959ha,Marciano:1976jc,Goldman:1977gh}.
The four-fermion interaction \eqref{4-fermion} changes the prediction
\begin{equation*}
  R = R_0\frac{|1+\Delta_P^e|^2}{|1+\Delta_P^\mu|^2} =
    R_0\frac{1+2\Real[\Delta_P^e]+\l\Real[\Delta_P^e]\r^2+
         \l\Imag[\Delta_P^e]\r^2}
       {1+2\Real[\Delta_P^\mu]+\l\Real[\Delta_P^\mu]\r^2+
         \l\Imag[\Delta_P^\mu]\r^2}\;.
\end{equation*}
Thus to the second order in $\Delta_P$ one obtains a constraint
\begin{gather}
  -2.9\times10^{-3} < f(\Delta_P^e,\Delta_P^\mu) < 0.4\times10^{-3}
  \label{bound} \\
\begin{split}
  f(\Delta_P^e,\Delta_P^\mu)=\Real[\Delta_P^e-\Delta_P^\mu]
  +\frac{1}{2}\Real[\Delta_P^e-\Delta_P^\mu]\Real[\Delta_P^e-3\Delta_P^\mu]+
  \\
  +\frac{1}{2}\Imag[\Delta_P^e-\Delta_P^\mu]\Imag[\Delta_P^e+\Delta_P^\mu]
+{\cal O}(\Delta^3)
\end{split}\nonumber
\end{gather}
at 95\% C.L\@.  Constraints on T-violating
correlations in $\pi_{l2\gamma}$ decays,
which result from Eq.~\eqref{bound}, are
model-dependent. Any constraints on the coupling 
constants $G_P^\mu$ and
$G_P^e$ following from Eq.~(\ref{bound}) are evaded, if there is a
hierarchy in the coupling constants, 
\begin{equation}\label{muehierarchi}
  G_P^\mu/G_P^e=m_\mu/m_e\;.
\end{equation} 
Then the contributions to $R$ cancel: 
indeed, if $\mu-e$ hierarchy \eqref{muehierarchi} is satisfied, then 
$\Delta_P^e=\Delta_P^\mu$ and the result for the ratio $R$ 
coincides with the SM prediction.  In this case any $\Imag[\Delta_P]$
are allowed, and though the decays we discuss are rare processes, 
Br$(\pi\to e\bar{\nu}_e\gamma)=
(1.61\pm0.23)\times10^{-7}$~\cite{Bolotov:1990yq,pdg2000}, 
Br$(\pi\to \mu\bar{\nu}_\mu\gamma)=
(2.00\pm0.25)\times10^{-4}$~\cite{Bressi:1998gs}, 
even experiments with relatively low pion statistics
have chances to observe T-violation in $\pi_{l2\gamma}$ decays: the
total number of charged pions should be $N_\pi\gtrsim10^8$ for 
$\pi_{\e2\gamma}$ and $N_\pi\gtrsim10^5$ for $\pi_{\mu2\gamma}$.  

Note that to the leading order in $\Delta_P$, the bound~(\ref{bound})
constrains only the real parts of the coupling constants $G_P^\mu$ and
$G_P^e$ entering Eq.~(\ref{4-fermion}),
while constraints on imaginary parts are
weaker.  Thus, for general $G_P^\mu$ and $G_P^e$ 
(if the hierarchy~(\ref{muehierarchi})
does not hold, i.e., if there is no cancellation between $\Delta_P^e$ and
$\Delta_P^\mu$) one obtains $|\Real[\Delta_P]|\lesssim10^{-3}$ and 
$|\Imag[\Delta_P]|\lesssim0.03$.  Hence in this
case experiments
aimed at searching for T-violation in $\pi_{l2\gamma}$ decays should
have sufficiently large statistics: the total number of charged pions
should be $N_\pi\gtrsim10^{11}$ for $\pi_{\e2\gamma}$ and
$N_\pi\gtrsim10^8$ for $\pi_{\mu2\gamma}$.  Note that in the
$\pi_{\mu2\gamma}$
case the contribution of the new interaction~(\ref{4-fermion}) to
T-odd correlation is at best of the same order of magnitude as FSI effects.  One
can hope to discriminate between them only if they have different signs,
i.e.\ $\Imag[\Delta_P^\mu]$ is negative (cf.  Figures 1 and 3).

In models with $\Real[G_P]\sim\Imag[G_P]$ and without
hierarchy~(\ref{muehierarchi}), the bound (\ref{bound}) from $\pi\to l\nu$
decays implies $|\Imag[\Delta_P]|\lesssim10^{-3}$, which
significantly constrains possible contribution of the new
interaction~(\ref{4-fermion}) to T-odd correlation in $\pi_{l2\gamma}$
decays.  Namely, the contribution to the $\pi_{e2\gamma}$ decay should
be of the same order or weaker than one from the Standard Model
FSI\@.  The constraint on $\Delta_P^\mu$ means that the effect is
at least
two orders of magnitude smaller than the FSI effects in
$\pi_{\mu2\gamma}$ decay.  Nevertheless, as we discussed,
in the case of
$\pi_{e2\gamma}$ the difference in ($x,y$) distributions of FSI and
four-fermion contributions may allow one to discriminate between the
two if they are of the same order of magnitude, and even if
the contribution of
four-fermion interaction~(\ref{4-fermion}) is somewhat weaker.  On
the other hand, the experiment aimed at probing T-violation in
$\pi_{e2\gamma}$ decay has to deal with very high pion statistics.
Indeed, to test four-fermion
interaction~(\ref{4-fermion}) at the level allowed by $\pi\to e\nu_e$,
i.e., at the level of $10^{-3}$, one has to collect not less than
$10^{13}$ charged pions, assuming
statistical uncertainty only.  For
$\pi_{\mu2\gamma}$ decay, even though the branching ratio is not very
small, the ($x,y$) distributions of FSI and
four-fermion contributions have similar shapes, while the expected
effect is at least two orders of magnitude smaller than the FSI effects in
$\pi_{\mu2\gamma}$ decay, making it hardly
possible to detect new sources of T-violation
in models without hierarchy \eqref{muehierarchi} and with
$\Real[G_P^\mu]\sim\Imag[G_P^\mu]$.  

Overall, in the case of the hierarchy \eqref{muehierarchi}, decays
$\pi\to l\nu$ do not constrain new T-violating interactions which
can be searched for in relatively low statistics experiments,
$N_\pi\gtrsim10^8$ for $\pi_{e2\gamma}$ and $N_\pi\gtrsim10^5$ for
$\pi_{\mu2\gamma}$.  In the worst case of no hierarchy
\eqref{muehierarchi} and $\Real G_P\sim\Imag G_P$, new T-violating
interactions have little chance to be observed, and in $\pi_{e2\gamma}$
decay only.

\section{A simple model: heavy pseudoscalar exchange}
\label{sec:3}

To illustrate that the hierarchy~(\ref{muehierarchi}) may appear
naturally in low-energy physics (thus the T-violating effects may be
sufficiently large) we present below an example of a model which can
lead to the effective interaction~(\ref{4-fermion}) with coupling
constants obeying the hierarchy~(\ref{muehierarchi}).  As we show, this
model may be considered as ``proof by example'' of the fact that in
extensions of the Standard Model, large T-violating low-energy
interaction~\eqref{4-fermion} may arise without any contradiction to
existing experiments.

Let us assume that in addition to SM content there exists a heavy
charged pseudoscalar field coupled to both lepton and quark sectors
via the following Lagrangian
\begin{equation}\label{Lhiggs}
  \cL_H = Y_{ij} H^* \bar{d_i}\gamma_5 u_j
        + Y_e H \bar{\nu}_e(1+\gamma_5)e
        + Y_\mu H \bar{\nu}_\mu(1+\gamma_5)\mu
        + \mathrm{h.c.}
\end{equation}
The mass of the charged ``extra Higgs'' particle $H$ is supposed to be
of the order of $M_H\sim 1$~TeV. We assume for simplicity that there 
are no mixing couplings in the
 leptonic sector.  The new
Yukawa coupling constants in the leptonic sector are supposed to obey
the hierarchy $Y_\mu/Y_e=m_\mu/m_e$ with accuracy of $1\div0.1$\%
at the scale $M_H$.  Then even for the model with
$\Real[Y]\sim\Imag[Y]$, the bound from $\pi\to l\nu$
decays~(\ref{bound}) gives no constraints on the contribution of the new
interaction~(\ref{Lhiggs}) to T-odd correlations in $\pi_{l2\gamma}$
decays.  

At energies below $M_H$ this Lagrangian leads to the following
four-fermion interaction
\begin{equation}\label{4fermion}
\begin{split}
  \cL_Y &= \Bigl(G^e_{P\,ij} \bar{d_i}\gamma_5 u_j\cdot\bar{\nu}_e(1+\gamma_5)e
    + G^\mu_{P\,ij} \bar{d_i}\gamma_5 u_j\cdot\bar{\nu}_\mu(1+\gamma_5)\mu
  \\
  & + G_P^{\mu e} \bar{\nu}_e(1+\gamma_5)e
             \cdot\bar{\mu}(1-\gamma_5)\nu_\mu + \mathrm{h.c.}\Bigr)
  - G_{P\,ijkl} \bar{u}_i\gamma_5 d_j \cdot \bar{d_k}\gamma_5 u_l\\
&
  +G_P^{ee}\bar{\nu}_e(1-\gamma_5)e\cdot\bar{e}(1+\gamma_5)\nu_e+
G_P^{\mu\mu}\bar{\nu}_\mu(1-\gamma_5)\mu\cdot\bar{\mu}(1+\gamma_5)\nu_\mu
  \;,
\end{split}
\end{equation}
where
\begin{align} \nonumber
  G^e_{P\,ij} &= \frac{Y_{ij}Y_e}{M^2} \;,
  &
  G^\mu_{P\,ij} &= \frac{Y_{ij}Y_\mu}{M^2} \;,
  &
  G_{P\,ijkl} &= \frac{Y_{ji}^*Y_{kl}}{M^2} \;,
\\ 
\label{GPdef}
  G^{\mu e}_P &= \frac{Y_\mu^* Y_e}{M^2} \;,
  &
  G^{e e}_P &= \frac{Y_e^* Y_e}{M^2} \;,
  &
  G^{\mu \mu}_P &= \frac{Y_\mu^* Y_\mu}{M^2} \;.
\end{align}
The $\mu-e$ hierarchy is natural in this model, since the
corresponding couplings $G_P^e\equiv G^e_{P\,11}$ and $G_P^\mu\equiv
G^\mu_{P\,11}$ emerge as a result of Yukawa (Higgs-like) interaction
with heavy charged scalar.  Assuming $\mu-e$ hierarchy we obtain
\begin{align}\label{deltap_e}
  \Imag[\Delta_P^e]&=500\cdot \Imag[Y_{ud}Y_e]\cdot 
  \l\frac{1~{\rm TeV}}{ M}\r^2=2.5\cdot \Imag[Y_{ud}Y_\mu]\cdot 
  \l\frac{1~{\rm TeV}}{ M}\r^2\,, \\\label{deltap_mu}
  \Imag[\Delta_P^\mu]&=500\cdot \Imag[Y_{ud}Y_\mu]\cdot 
  \l\frac{1~{\rm TeV}}{ M}\r^2\,.
\end{align}
Though $\Imag[\Delta_P^\mu]$ is significantly larger than
$\Imag[\Delta_P^e]$, the resulting contributions to the T-odd
correlations are of the same order.  Indeed, the asymmetry $P_T$ is
determined by the product of $\Imag[\Delta_P]$ and the distribution
function $[\sigma_V(x,y)-\sigma_A(x,y)]$ (see Eq.~(\ref{pt})).  The
typical values of this function for $\pi_{2e\gamma}$ and
$\pi_{2\mu\gamma}$ decays (see Figure 1) exhibit the hierarchy inverse
to one appearing in Eqs.~(\ref{deltap_e}), (\ref{deltap_mu}).  Thus
one concludes that the large observable T-violating effect (of the
order of $1\div0.1$) in both $\pi_{e2\gamma}$ and $\pi_{\mu2\gamma}$
is possible if $\Imag[Y_{ud}Y_\mu]\simeq1\div0.1$.

In the following subsections we discuss the experimental constraints on the
interaction~(\ref{Lhiggs}).

\subsection{Muon decay}
\label{sec:3.1}

If $\mu-e$ hierarchy \eqref{muehierarchi} holds, then the interaction
\eqref{Lhiggs} contributes to $\mu\to e\bar{\nu}_e\nu_\mu$ decay.  The
Standard Model contribution to the squared modulus of the matrix
element summed over spin states
\begin{equation*}
  \overline{|M_{SM}|^2} = 128G_F^2(pq_1)(kq_2)
\end{equation*}
has to be compared with the contribution from the interaction
\eqref{4fermion},
\begin{equation*}
  \overline{|M_P|^2} = 64|G_P^{\mu e}|^2(pq_2)(kq_1) \;,
\end{equation*}
and with the interference contribution
\begin{equation*}
  2M_{SM}\Real[M_P] = 
  128\Real[G_P^{\mu e}]\frac{G_F}{\sqrt{2}}m_\mu m_e (q_1q_2) \;.
\end{equation*}
Here $p$, $k$, $q_1$, $q_2$ are momenta of $\mu$, $e$, $\bar{\nu}_e$,
and $\nu_\mu$, respectively.  Integrating over neutrino momenta one
obtains for the differential muon decay width
\begin{equation}
\label{muonwidth}
d\Gamma =\frac{(G_F^2+|G_P^{\mu e}|^2/2)m_\mu^5}{96\pi^3}
     (3-2\varepsilon)\varepsilon^2d\varepsilon
   +\frac{\Real[G_P^{\mu e}]G_Fm_em_\mu^4}{8\sqrt{2}\pi^3}
    (1-\varepsilon)\varepsilon d\varepsilon \;,
\end{equation}
where $\varepsilon=k^0/k^0_{\mathrm{max}}=2k^0/m_\mu$.

Experimental constraint on the coupling constant $G^{\mu e}_P$
entering this formula can be obtained from the data on the $\rho_0$
parameter \cite{pdg2000,Erler:1998df},
\begin{equation*}
  \rho_0 = \frac{M_W^2}{M_Z^2\hat{c}_Z^2\hat{\rho}(m_t,M_H)} \;,
\end{equation*}
which is the measure of the neutral to charged current interaction
strength (here $\hat{c}_Z$ is cosine of Cabibbo angle in
$\overline{\mathrm{MS}}$ scheme, and $\hat{\rho}$ absorbs the SM
radiative corrections).  At one sigma level this parameter
equals~\cite{pdg2000}
\begin{equation*}
  \rho_0 = 0.9998_{-0.0006}^{+0.0011} \;.
\end{equation*}
If we postulate that neutral current interactions are precisely the
same as in the SM, then we allow new physics contribution to the muon
decay width of the order of
\begin{equation*}
  |\Delta\Gamma| \lesssim 2(1-\rho_0) \Gamma_{SM}
\end{equation*}
(the factor 2 appears here because $\Gamma\sim G_F^2\sim M_W^{-4}$).
Comparing this deviation with the term proportional to $|G_P^{\mu
e}|^2$ in Eq.~\eqref{muonwidth} we obtain
\begin{equation}
|G_P^{\mu e}|<0.07G_F\;.
\label{muon-bound}
\end{equation}
In fact, this is a fairly weak constraint.  For example, setting
$Y_\mu=1$ and assuming $\mu-e$ hierarchy, one obtains from
Eq.~(\ref{muon-bound})
\begin{equation*}
  M > Y_\mu^2 \sqrt{\frac{m_e}{m_\mu}\frac{1}{0.07G_F}}
    \simeq 75~{\rm GeV} \;.
\end{equation*}
The account of the second term in \eqref{muonwidth} leads to even
weaker constraint on $\Real[G_P^{\mu e}]$.  So, constraints on the
leptonic part of the Lagrangian \eqref{Lhiggs} are very modest.

\subsection{$K^0-\overline{K^0}$ mixing}
\label{sec:3.2}

\begin{figure}[htb]
  \begin{center}
    \includegraphics{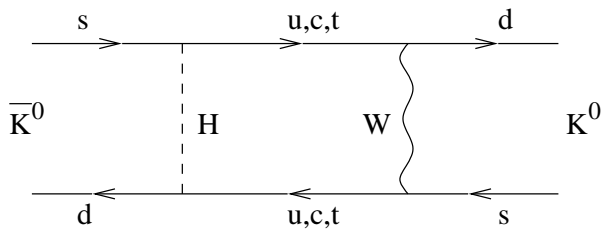}
    \caption{Box diagram contributing to $K^0$--$\overline{K^0}$ 
      mass difference.
      \label{fig:box}}
  \end{center}
\end{figure}

Let us estimate the contribution from the lagrangian~(\ref{Lhiggs}) 
to the mixing parameters in neutral kaon system.  These are $K_L-K_S$ 
mass difference $\Delta m$ and CP-violating parameter $\epsilon$.  
They are expressed as follows,
\begin{equation*}
  \Delta m =
  \frac{2\Real\bra{K^0}H_{\mathrm{eff}}\ket{\overline{K^0}}}{2m_K}\;,~~~
\epsilon = 
    \frac{\Imag\bra{K^0}H_{\mathrm{eff}}\ket{\overline{K^0}}}{2\Real\bra{K^0}
H^{\mathrm{full}}_{\mathrm{eff}}\ket{\overline{K^0}}}
  \;,
\end{equation*}
where $H_{\mathrm{eff}}$ is the effective four fermion Hamiltonian,
generated by the diagram of Figure~\ref{fig:box} and
$H_{\mathrm{eff}}^{\mathrm{full}}$ is the effective Hamiltonian
containing contributions both from the diagram presented in
Figure~\ref{fig:box} and from the usual box diagram with two $W$
bosons.  We obtain
\begin{align}
  H_{\mathrm{eff}}
  &= \frac{g^2}{8}\sum_{i,j=u,c,t} U_{is}U_{jd}^*Y_{js}Y_{id}^*
     \big\{[4\bar{d}_L\gamma^\mu s_L \cdot
          \bar{d}_L\gamma^\mu s_L]
          I_2(m_i,m_j)  \nonumber\\
  &\quad +[16\bar{d}_Rs_L\cdot\bar{d}_Ls_R
         -4\bar{d}_R\sigma^{\rho\mu}s_L \cdot
           \bar{d}_L\sigma^{\rho\mu}s_R ]
          I_1(m_i,m_j) 
  \big\}
\label{heff} 
\end{align}
where $U$ is the Cabibbo--Kobayashi--Maskawa mixing matrix,
$q_{R,L}=\frac{1\pm\gamma^5}{2}q$ and
\begin{align*}
  g_{\rho\lambda}I_1(m_i,m_j) & = i\int \frac{d^4q}{(2\pi)^4}
    \frac{q_\rho q_\lambda}{(q^2-m_i^2)(q^2-m_j^2)(q^2-M_W^2)(q^2-M_H^2)} \\
             & = g_{\rho\lambda}\frac{1}{2(4\pi)^2M_H^2}
                  A_1(\frac{m_i}{M_H},\frac{m_j}{M_H},\frac{M_W}{M_H})
  \\
  I_2(m_i,m_j) & = i\int \frac{d^4q}{(2\pi)^4}
    \frac{m_im_j}{(q^2-m_i^2)(q^2-m_j^2)(q^2-M_W^2)(q^2-M_H^2)} \\
             & = -\frac{m_im_j}{(4\pi)^2M_H^4}
                  A_2(\frac{m_i}{M_H},\frac{m_j}{M_H},\frac{M_W}{M_H})
\end{align*}
with
\[
  A_k(\alpha,\beta,\gamma)
  = \int\limits_0^1dx\,dy\,dz\,dw
    \frac{\delta(x+y+z+w-1)}{[\alpha^2x+\beta^2y+\gamma^2z+w]^k}\;,~~~k=1,2\;.  
\]  
We calculate the matrix elements of the operators in Eq.~\eqref{heff}
by making use of the ``vacuum saturation'' approximation, where
non-vanishing matrix elements are
\begin{align*}
  \bra{K^0}\bar{d}_L\gamma^\mu s_L\cdot\bar{d}_L\gamma^\mu s_L
\ket{\overline{K^0}}
  & = \frac{2}{3}m_K^2f_K^2\;, \\
  \bra{K^0}\bar{d}_Rs_L\cdot\bar{d}_Ls_R\ket{\overline{K^0}}
  & = \left[\frac{m_K^2}{12}+\frac{B_0^2}{2}\right]f_K^2
\end{align*}
Therefore
\begin{multline}\label{mkkbar}
  \bra{K^0}H_{\mathrm{eff}}\ket{\overline{K^0}} =
  \frac{g^2f_K^2}{3(4\pi)^2}
  \sum_{i,j=u,c,t} U_{is}U_{jd}^*Y_{js}Y_{id}^*
   \times\\
  \times \left\{\frac{m_K^2+6B_0^2}{4M_H^2}
                  A_1(\frac{m_i}{M_H},\frac{m_j}{M_H},\frac{M_W}{M_H})
  \right.
  \\
  \left.
  -\frac{m_im_jm_K^2}{M_H^4}
                  A_2(\frac{m_i}{M_H},\frac{m_j}{M_H},\frac{M_W}{M_H})
    \right\}\;.
\end{multline}
For $M_H\sim1\mathrm{TeV}$ the second term in parenthesis is
negligible even for $t$-quark running in the loop.  Thus we obtain  
\begin{align}\label{finalmkkbar}
  \Delta m  &=
  \frac{g^2f_K^2(m_K^2+6B_0^2)}{12(4\pi)^2M_H^2m_K}\times\\
&\quad\times
  \sum_{i,j=u,c,t} \Real[U_{is}U_{jd}^*Y_{js}Y_{id}^*]
   \times A_1(\frac{m_i}{M_H},\frac{m_j}{M_H},\frac{M_W}{M_H})\;,
\nonumber\\
\epsilon  &=
  \frac{g^2f_K^2(m_K^2+6B_0^2)}{24(4\pi)^2M_H^2}\frac{1}{\Real\bra{K^0}
H^{\mathrm{full}}_{\mathrm{eff}}\ket{\overline{K^0}}}\times
\label{finaleps}
\\
&\quad\times
  \sum_{i,j=u,c,t} \Imag[U_{is}U_{jd}^*Y_{js}Y_{id}^*]
   \times A_1(\frac{m_i}{M_H},\frac{m_j}{M_H},\frac{M_W}{M_H})\;.
\nonumber
\end{align}
The contribution to the $K^0-\overline{K^0}$ mass difference
\eqref{mkkbar} should be smaller than the experimental value $\Delta
m=3.489\times10^{-12}$MeV and CP-violating parameter $\epsilon$ should
not exceed the experimental value $2.271\times10^{-3}$, which are both
consistently described by the CP violating CKM matrix in the Standard
Model.  This constrains the coupling constants $Y_{ij}$.

The strongest constraint comes from the measurement of CP-violating
parameter $\epsilon$, while, generally, the constraint from the
$K_L-K_S$ mass difference is an order of magnitude weaker.  If one
assumes that all coupling constants $Y_{ij}$ are of the same order and
have complex phases of order one, this measurement requires them to be
less than $10^{-4}$, and if we set $Y_\mu=1$ then one obtains
$P_T\sim10^{-4}$ (see Eqs.~(\ref{deltap_e}), (\ref{deltap_mu}) and
Fig.~\ref{fig:sigma}).  Obviously, the situation changes for
nontrivial structure of the $Y$ matrix.  If the coupling with the
first generation quarks is larger than with the second generation,
then $Y_{ud}$ (and, correspondingly, $\Delta_P$) can be quite large,
even if $\epsilon$ is kept small.  As an example the values
$|Y_{ud}=0.01|$, $|Y_{us}|=|Y_{cd}|=10^{-5}$, $|Y_{cs}|=10^{-4}$ are
allowed.  Another promising pattern is $Y_{ij}\propto U_{ij}$, i.e.
coupling constants $Y$ are proportional to CKM matrix.  This is the
case if $Y$-matrix is proportional to the unit matrix in gauge basis
of quarks.  The constraint disappears also in the case of aligned
complex phases of $Y_{ij}$; in that case, the constraint from kaon
mass difference becomes important.  And, of course, if $Y_{is}$ is
zero, then there is no contribution to $\epsilon$ and $\Delta_m$ at
all.

Note that measurement of parameter $\epsilon$ gives the strongest
limit on our model in comparison with other CP-violating effects in
kaon physics.  Contribution to $\epsilon'/\epsilon$ may be estimated
following the lines of Ref.~\cite{Barger:1990fj}.  It is negligible
for any parameters allowed by the requirement
$\epsilon<\epsilon_{SM}$.  Transverse muon polarization in
$K_{\mu2\gamma}$ decay is relevant only for the special choice of
$Y\sim U$, since in this case the interaction \ref{4fermion} leads to
$P_T$ in this process of the same order as in $\pi_{e2\gamma}$.  For
other choices of $Y$ transverse muon polarization in $K_{\mu2\gamma}$
is much smaller than in pion decays.

\subsection{Neutron dipole moment}
\label{sec:3.3}

\begin{figure}[htb]
  \begin{center}
    \includegraphics{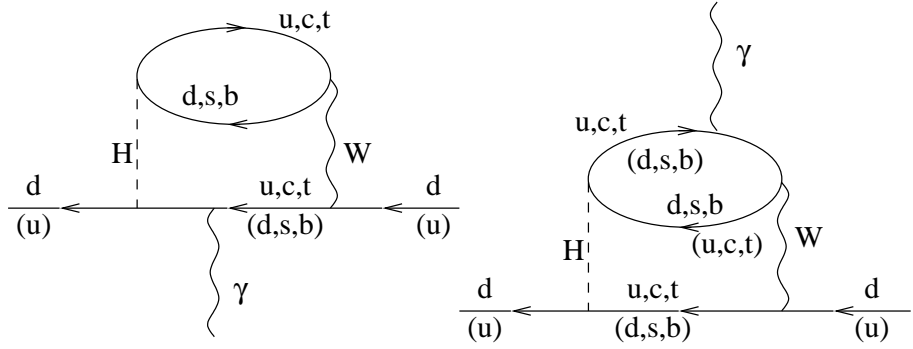}
    \caption{Typical two-loop diagrams contributing to 
             neutron dipole moment $d_n$.
      \label{fig:two-loop}}
  \end{center}
\end{figure}

In our model, heavy charged scalar particles give a contribution to
neutron dipole moment at two-loop level (see
Figure~\ref{fig:two-loop}).
The contribution of the diagrams in Figure~\ref{fig:two-loop} can be
estimated as
\begin{equation}\label{dipole-estimate}
\begin{split}
  \frac{d_n}{e} \sim &\frac{\left(\frac{g}{2\sqrt{2}}\right)^2}{16\pi^2}
  \sum_{i,j,k}\frac{Y_{dj}U_{jd}Y_{ik}^*U_{ki}^*}{16\pi^2}\times\\\times & 
  \frac{m_i f\l\frac{m^2_i}{M_H^2},\frac{m^2_k}{M_H^2},
   \frac{m_j^2}{M_H^2},\frac{M_W^2}{M_H^2}\r
+m_k g\l\frac{m^2_i}{M_H^2},\frac{m^2_k}{M_H^2},
   \frac{m_j^2}{M_H^2},\frac{M_W^2}{M_H^2}\r}{M_H^2}
  \;,
\end{split}
\end{equation}
where dimensionless functions $f$ and $g$ are of order one.  For
$m_i=10$MeV, $M_H\simeq1$TeV and $Y_{ij}\simeq1$ we have
$d_n/e\sim10^{-27}$cm, which is two orders of magnitude smaller than
the current limit.  For the second generation of quarks running in the
loop we have $m_i\lesssim1$GeV, which still allows corresponding
Yukawas to be of order $1$.  All special forms of $Y$ matrix described
in the previous subsection lead to acceptable contribution to neutron
dipole moment.  Thus, the constraints on the parameters of the
lagrangian \eqref{Lhiggs} from $K-\overline{K}$ mass difference are
more stringent than current bounds from neutron dipole moment
measurements.

To summarize the results of this section, the new interactions
\eqref{Lhiggs} cannot lead to significant T-violating effects in
$\pi_{l2\gamma}$ decays if the Yukawa couplings do not exhibit a
special pattern.  In that case the most stringent constraints come
from the measurement of $\epsilon$-parameter.  However, there are a
number of special cases when these and other constraints do not apply,
and T-violation in $\pi_{l2\gamma}$ decays is allowed to be
sufficiently large.  Hence the lepton transverse polarization
asymmetry in $\pi_{l2\gamma}$ is an interesting probe of the structure
of new interactions at TeV scale.

\section{Acknowledgments}

\begin{sloppypar}
The authors are indebted to V.~N.~Bolotov, Yu.~G.~Kudenko and
V.~A.~Rubakov for stimulating discussions.  The work is supported in
part by CPG and SSLSS grant 00-15-96626, by RFBR grant 02-02-17398 and
by the program SCOPES of the Swiss National Science Foundation,
project No.~7SUPJ062239.  The work of F.B.\ is supported in part
by CRDF grant RP1-2364-MO-02.  The work of D.G.\ is supported in
part by the RFBR grant 01-02-16710 and INTAS YSF 2001/2-142.

\end{sloppypar}


\end{document}